\documentclass[useAMS,usenatbib,a4paper]{mn2e}
\usepackage{graphicx}
\usepackage{amsmath}
\usepackage{amssymb}

 \title[Cold gas in the Intra
Cluster Medium]{Cold gas in the Intra Cluster Medium: implications for
flow dynamics and powering optical nebulae}

\author[E.C.D. Pope, T.W. Hartquist, J.M. Pittard] {Edward
C.D. Pope\thanks{E-mail:e.c.d.pope@leeds.ac.uk}, Thomas W. Hartquist,
Julian M. Pittard\\ School of Physics \& Astronomy, University of
Leeds, Leeds, UK, LS2 9JT\\ }

\begin{document}

\pagerange{\pageref{firstpage}--\pageref{lastpage} \pubyear{2008}}

 \maketitle

\label{firstpage}

\begin{abstract}
We show that the mechanical energy injection rate generated as the
intra-cluster medium (ICM) flows around cold clouds may be sufficient
to power the optical and near infra-red emission of nebulae observed
in the central regions of a sample of seven galaxy clusters.  The
energy injection rate is extremely sensitive to the velocity
difference between the ICM and cold clouds, which may help to explain
why optical and infra-red luminosity is often larger than expected in
systems containing AGNs. We also find that mass recycling is likely to
be important for the dynamics of the ICM. This effect will be
strongest in the central regions of clusters where there is more than
enough cold gas for its evaporation to contribute significantly to the
density of the hot phase.
\end{abstract}


\section{Introduction}

Line-emitting nebulae surround approximately a third of all Brightest
Cluster Galaxies (BCGs) \citep[][]{crawf}. The occurance of these
optical nebulae appears to have some correspondence to the short
radiative cooling time of the cluster. Nearby examples, such as
NGC1275, in Perseus, NGC4696 in Centaurus, and A1795, show extended
filamentary structures up to 50 kpc from the central galaxy
\citep[see][and references therein]{hatch07}. Many of these BCGs also
contain reservoirs of $10^{8}-10^{11.5}\,{\rm M_{\odot}}$ of molecular
hydrogen \citep[e.g.][]{edge01}.

One of the prevalent emission lines in these nebulae is [NII]$\lambda
6584$. Since [NII] is collisionally excited by thermal electrons, this
line measures the heating rate in the gas, whereas the H$\alpha$ line
measures the ionisation rate \citep[][]{don91}. A large heating rate
compared to the ionisation rate leads to large [NII]/H$\alpha$ and
[OII]/H$\alpha$ ratios. To date, the power source remains unknown,
although numerous heating processes have been investigated. These
include ionisation by the central active galactic nucleus (AGN),
ionisation by hot young stars, heating by X-rays from the ICM,
conduction of heat from the ICM \citep[][]{don00}, and turbulent
mixing layers \citep[][]{crawf92}. Cosmic rays, preferentially
diffusing along the magnetic field lines trailing behind rising
bubbles, could possibly drive the excitation in those filaments that
are located in bubble wakes \citep[][]{rusz07}. However, there are
problems associated with each possiblity listed above
\citep[see][]{hatch}. As a consequence, the nature of the power
source(s) remains unknown.

Energy dissipation due to drag has, so far, not been considered as a
potential source of energy for the filaments. In principle,
dissipation by drag could be responsible for at least the heating of
the optically emitting gas, and thus the [NII] emission, if not the
H$\alpha$ emission.

Recent numerical simulations suggest that the filamentary structures
can be formed by an outflow stripping material from cold clouds
\citep[][]{pope08}. These filaments only become long if the wind is
sufficiently fast. The kinematics and morphology also suggest that the
filaments may be many tens of Myr old. Dissipation due to drag would
necessarily occur all along the filament and would provide a spatially
distributed source of energy that could power the optical emission.

So far the presence of vast quantities of cold material has been
ignored in most numerical simulations of the diffuse gas in galaxy
clusters. This is partly due to the difficulty of implementing a cold
phase in such simulations, but also because the importance of its
presence has not been hitherto established. The effect of a cold phase
may not be entirely negligible. In fact, the presence of significant
quantities of cold gas might alter the flow dynamics in the central
regions of galaxy clusters. This could occur through the addition of
mass, stripped from cold clouds, or energy and momentum transfer
associated with the mass loss from the clouds.

In this article we estimate mass transfer parameters for a sample of
galaxy clusters: Virgo, Perseus, Hydra, A2597, A2199, A1795 and
A478. We also investigate drag as a possible power source for the
optical emission of these nebulae. In section 2 we discuss the
constraints on cloud properties. In section 3 we concentrate on the
significance of mass transfer between the phases. Section 4 focusses
on mechanical energy as a power source for the optical emission, and
we summarise in section 5.

\section{Constraints on cold clouds}

There is little direct evidence constraining cloud sizes in the
ICM. For example, if the optical filaments are drawn from cold clouds
the filaments are likely to be no wider than these clouds. However,
the width of the filaments are usually constrained to be $<$ 1 kpc
which is not a very strong limit. Maps presented by \cite{salome08}
show clumpy structures in the CO emission. Given the spatial
resolution of the technique, the upper limit on the sizes of these
clumps is $\sim 450\,{\rm pc}$ in the Perseus cluster. Such clumps may
be complexes of giant molecular clouds, but it is difficult to tell.

In the simplest cloud formation scenario there is a heat conduction
length scale above which fluctuations are not stabilised and clouds
form, which is given by \cite{boeh89} \citep[following][]{field} as
\begin{equation}\label{eq:cloudrad}
\lambda > 6
\bigg(\frac{\eta}{0.01}\bigg)^{1/2}\bigg(\frac{T}{10^{8}\,{\rm
K}}\bigg)^{3/2}\bigg(\frac{n}{0.1\,{\rm cm^{-3}}}\bigg)^{-1} \,{\rm
kpc},
\end{equation}
where $\eta$ is the thermal conduction suppression factor, $T$ is the
ambient temperature, and $n$ is the number density of the
gas. Equation (\ref{eq:cloudrad}) implies a minimum cloud mass of
\begin{equation}\label{eq:mcrit}
M > 1.7 \times 10^{8}
\bigg(\frac{\eta}{0.01}\bigg)^{3/2}\bigg(\frac{T}{10^{8}\,{\rm
K}}\bigg)^{9/2}\bigg(\frac{n}{0.1\,{\rm cm^{-3}}}\bigg)^{-2} \,{\rm
M_{\odot}}.
\end{equation}
\cite{mc77} present a similar argument. Equation (\ref{eq:mcrit})
indicates that smaller, lower mass clouds can form near the central
regions of galaxy clusters, where the gas density is at its
highest. More massive clouds may form further out, where the density
is lower.

However, the physics of the ICM is rather more complex than this
description implies. Processes such as magnetic turbulence are likely
to play roles, with $\eta$ possibly as small as
$10^{-6}$. Consequently, clouds much smaller in mass than
$10^{8}\,{\rm M_\odot}$ may form. Given the difficulty of observing
single cold clouds, much of the work on cloud properties has relied on
theoretical reasoning \citep[e.g.][]{daines,loew}. However, there
remains no clear picture of the mass and size of individual
clouds. All we can say for certain is that, based on filament widths,
and the CO data, the clouds are probably less than 0.5 kpc wide and
there is likely to be a distribution of cloud sizes. Also, the cold
mass should not be confused with the mass of optically-emitting
ionised material, which can be determined from the H$\alpha$ emission
and is much less than the total quantity of cold gas.

If the most massive clouds are molecular, much of the cloud material
may be at temperatures of tens of Kelvin. Pressure equilibrium with
the surrounding X-ray emitting gas would imply a density of order
$10^{-20}-10^{-19}\,{\rm g\,cm^{-3}}$. A spherical cloud of
$10^{8}\,{\rm M_\odot}$ would have a diameter of $\sim 100\,{\rm
pc}$. Though the cloud mass is very uncertain, we will typically adopt
values of $10^{-19}\,{\rm g\,cm^{-3}}$, $100\,{\rm pc}$ and
$10^{8}\,{\rm M_\odot}$ for the density, diameter and mass where
presenting numerical results. However, we will make the ways in which
our results scale with the assumed values very clear.

\subsection{The motion of clouds}

The clouds will be referred to as the dispersed phase, while the
intracluster gas will be referred to as the continuous phase denoted
by subscripts $\rm d$ and $\rm c$, respectively.

Concentrating only on drag, we can write the equation of motion of a
cold cloud through the ICM as
\begin{equation} \label{eq:mot}
m\frac{{\rm d}v}{{\rm d}t} = \frac{1}{2}C_{\rm D}\frac{\pi
D^{2}}{4}\rho_{\rm c}(u-v)|u-v|,
\end{equation}
where $C_{\rm D}$ is the drag coefficient, $D$ is the diameter of the
cloud, $\rho_{\rm c}$ is the density of the continuous phase, $u$ is
the speed of the continuous phase, and $v$ is the speed of the
dispersed phase.

Dividing both sides of equation (\ref{eq:mot}) by the cloud mass and
assuming that the cloud is spherical with uniform density, we find
\begin{equation} \label{eq:mot2}
\frac{{\rm d}v}{{\rm d}t} = \frac{(u-v)}{\tau_{\rm v}},
\end{equation}
where
\begin{equation}\label{eq:mot3}
\tau_{\rm v} = \frac{4D}{3C_{\rm D}}\bigg(\frac{\rho_{\rm
d}}{\rho_{\rm c}}\bigg)\frac{1}{|u-v|}.
\end{equation}
Equations (\ref{eq:mot2}) and (\ref{eq:mot3}) mean that the cloud's
motion can be described in terms of the drag coefficient. The drag
coefficient is a function of the Reynolds number of the fluid flowing
around an obstacle, and the Mach number (although for subsonic flows
this effect is small). The Reynolds number is given by
\begin{equation}
Re = \frac{D \rho_{\rm c}|u-v|}{\mu_{\rm c}},
\end{equation}
where $\mu_{\rm c}$ is the viscosity of the continuous phase. An
approximate relation for the subsonic drag coefficient for flow past a
solid object at $Re < 1000$ is given by \cite{multi},
\begin{equation} \label{eq:cd}
C_{\rm D} \approx \frac{24}{Re} + \frac{4}{Re^{2/3}}.
\end{equation}
So, $Re = 1$ gives $C_{\rm D} = 28$. At high Reynolds numbers ($Re >
1000$) the drag coefficient is approximately constant, $C_{\rm D}
\approx 0.45$. This is not true if the Mach number is $\ge 1$
\citep[e.g.][]{henderson}. In such a circumstance the drag coefficient
increases. However, the effect of the Mach number is minimal for
subsonic flows.

Typical cloud velocities can be estimated by various means. In the
Perseus cluster, the properties of the cold gas were probed with
absorption of 21 cm radio emission from a background radio source
\citep[][]{jaffe90}. Typical linewidths found were $100-500\,{\rm
km\,s^{-1}}$, which greatly exceeds the expected thermal
width. Similar linewidths were also found by \cite{edge01} and
\cite{salome03} from CO emission, and also from molecular hydrogen
emission by \cite{jaffe97}. These linewidths are comparable with
inflow velocities in the ICM. However, this should not be taken as an
indication that the clouds must be approximately co-moving with the
ICM. 

The Stokes number is a useful quantity to consider. It is defined as
the ratio of the momentum transfer timescale of the particle,
$\tau_{\rm v}$, to the dynamic timescale of the fluid flow around the
particle. The Stokes number is
\begin{equation}\label{eq:stoke}
St = \frac{\tau_{\rm v}(u-v)}{D}.
\end{equation}
Substitution for $\tau_{\rm v}$ gives
\begin{equation}
St = \frac{4}{3}\bigg(\frac{\rho_{\rm d}}{\rho_{\rm
c}}\bigg)\bigg(\frac{1}{C_{\rm D}}\bigg),
\end{equation}
which is independent of the cloud size and the relative
velocity. Since $\rho_{\rm d}/\rho_{\rm c} \gg 1/C_{\rm D}$, we have
$St \gg 1$. In this limit, the particle motion is not well coupled to
that of the continuous phase and the two phases effectively move
independently. This means that cold molecular cloud and the ICM
velocities are not coupled by viscosity.

Subsonic bulk flows in the ICM are currently undetectable. Therefore,
we need an alternative method to estimate the quantity $u-v$ that is
important for our calculations. Thus, the \cite{pope08} estimate of
$|u-v|$ (derived from a model of the formation of filaments due to an
AGN driven outflow) may provide a reasonable estimate for the Perseus
cluster.

\subsection{Definitions}

Before calculating the mass transfer coefficients it is necessary to
define some useful parameters. The first of these is the
volume-averaged density of the dispersed phase (assuming identical
clouds)
\begin{equation} \label{eq:rhodbar}
\bar{\rho}_{\rm d} = n_{\rm d}m = \alpha_{\rm d}\rho_{\rm d},
\end{equation}
where $n_{\rm d}$ is the number density of clouds of volume filling
factor $\alpha_{\rm d}$ and mass density $\rho_{\rm d}$. Similarly for
the continuous phase we have
\begin{equation}
\bar{\rho}_{\rm c} = \alpha_{\rm c}\rho_{\rm c},
\end{equation}
and the density of the mixture can be written $\rho_{\rm m} =
\bar{\rho}_{\rm c} + \bar{\rho}_{\rm d}$.

However, for ease of comparing with observations, it makes sense to
write equation (\ref{eq:rhodbar}) in a different way. The
volume-averaged density is simply the total mass of the dispersed
phase divided by the total volume of the system
\begin{equation}
\bar{\rho}_{\rm d} = \frac{M_{\rm d}}{V},
\end{equation}
where $ V$ is the volume of a spherical shell, and $M_{\rm d}$ is the
cold mass within the shell. Similarly for the continuous phase we have
\begin{equation}
\bar{\rho}_{\rm c} = \frac{ M_{\rm c}}{V}.
\end{equation}

\section{Mass exchange}

It is possible to assess the extent of the coupling between the phases
by evaluating so-called `coupling' parameters \citep[][]{multi}. The
continuous phase is described by the density, the temperature and the
velocity field. The particle/dispersed phase is described by the
concentration of particles, their size, their temperature, and their
velocity field. Coupling can take place through mass, momentum and
energy transfer between phases. Mass coupling is the addition of mass
through evaporation, or the removal of mass from the continuous phase
through condensation. Momentum coupling is the result of the drag
force on the dispersed and continuous phases. It can also occur due to
mass transfer between the phases. Energy coupling occurs through heat
transfer between the phases as well as the dissipation of kinetic
energy due to drag. Thermal and kinetic energy can also be transferred
between the phases by mass transfer. A quick numerical exercise
demonstrates that estimates of the energy and momentum coupling,
except that associated with mass exchange, are generally subject to
large uncertainties, and therefore unreliable.

\subsection{Definition of the mass coupling parameter}

Suppose that there are $n_{\rm d}$ clouds per unit volume in a cubic
box with side $X$. If each cloud evaporates at a rate $\dot{m}$, the
rate at which mass is injected by the dispersed phase in this volume
is
\begin{equation}
\dot{M}_{\rm d} = n_{\rm d} X^{3}\dot{m} = N\dot{m}.
\end{equation}
The mass flux of the continuous phase through this volume is
\begin{equation}
\dot{M}_{\rm c} \sim \bar{\rho}_{\rm c}u X^{2},
\end{equation}
and the mass coupling parameter is defined as \citep[][]{multi}
\begin{equation} \label{eq:mcoup}
\Pi_{\rm mass} = \frac{\dot{M}_{\rm d}}{\dot{M}_{\rm c}}.
\end{equation}
If $\Pi_{\rm mass} \ll 1$, then the effect of mass addition to the
continuous phase would be insignificant. If $\Pi_{\rm mass} \gg 1$,
then the mass added to the system would dominate. The latter would be
unphysical in galaxy clusters if it occurred globally, since it would
imply that the total mass of gas was not conserved. However, in a
restricted region this scenario would be consistent with a fountain
flow, where the clouds are replenished over and over again. In the
case where $\Pi_{\rm mass} \sim 1$ both condensation and evaporation
are important.

The mass coupling parameter is likely to be a function of radius in
galaxy clusters. Therefore, equation (\ref{eq:mcoup}) is of little
practical use, since we cannot compare like with like: the mass flow
rate of the continuous phase must be evaluated at a single radius,
while the mass injected into the continuous phase corresponds to a
volume.

In galaxy clusters, $\dot{M}_{\rm d}(r)$ is the net rate of mass added
to the flow from the clouds (evaporation - condensation), while
$\dot{M}_{\rm c}(r)$ is the mass flux of the hot gas. Several
scenarios are possible. If there is no evaporation/condensation
(i.e. no clouds at all), then $\dot{M}_{\rm c} = \dot{M}_{\infty}$,
where $\dot{M}_{\infty}$ is the pure flow rate of hot material. The
accretion rate of new material onto the central galaxy, $\dot{M}_{\rm
gal}$, is given by the value of the pure flow rate at the
characteristic radius of the galaxy: $\dot{M}_{\infty}(r_{\rm
gal})$. If there is condensation of material out of the hot phase, but
no subsequent evaporation, then $\dot{M}_{\rm c} <
\dot{M}_{\infty}$. Of course, if the clouds also accrete onto the
central galaxy, $\dot{M}_{\rm gal}$ may still be equal to
$\dot{M}_{\infty}$. A third possibility is that material evaporated
from the clouds may contribute a significant fraction of $\dot{M}_{\rm
c}$, and so $\dot{M}_{\rm c}$ may exceed the actual accretion rate of
``new'' material from infinity (i.e. $\dot{M}_{\rm c} >
\dot{M}_{\infty}$). Such a scenario would require that the mass in the
clouds is predominantly from material that was previously accreted
onto the central galaxy, and which has at some later time been thrown
out of the central galaxy back into the ICM, perhaps through AGN
activity.

Rates of mass deposition from the continuous phase to the cloud phase
are inferred from X-ray data. The mass deposition in a typical cluster
is consistent with a mass flow rate that increases linearly with
radius, $\dot{M}_{\rm obs} = -\dot{M}_{0}(r/r_{\rm cool})$, at $r <
r_{\rm cool}$ \citep[e.g.][]{pope06}. Here $\dot{M}_{0}$ is a constant
and $r_{\rm cool}$ is the radius at which the ICM's radiative cooling
time equals the Hubble time. If no evaporation of clouds occurred it
would be reasonable to assume that the flow rate of hot gas at $r_{\rm
cool}$ is $\dot{M}_{\rm c} = \dot{M}_{0}$ and that the flow rate of
hot gas at $r<r_{\rm cool}$ is $\dot{M}_{\rm c} =\dot{M}_{0}r/r_{\rm
cool}$. Now imagine the scenario where for $r_{\rm d} < r < r_{\rm
cool}$ there is pure condensation, while for $r < r_{\rm d}$ material
may also evaporate from clouds back into the hot phase so that there
is a complex exchange of mass between the hot gas and cold clouds
(i.e. condensation and evaporation, and re-condensation and
re-evaporation). Much of the condensing gas at $r < r_{\rm d}$ may be
gas that previously evaporated from clouds recently expelled from the
central galaxy in a fountain flow, or via AGN activity, for example,
rather than material that fell from $r \gg r_{\rm d}$. Thus, the
instantaneous accretion rate onto the galaxy may exceed the long-term
average (i.e. $\dot{M}_{\rm gal} \ge \dot{M}_{\infty}(r_{\rm
gal})$). Therefore, the observationally inferred mass flow rate
$\dot{M}_0$ may actually be an upper bound on the pure mass flow rate,
$\dot{M}_{\infty}$, with the real flow rate of ``new'' material being
$\beta \dot{M}_0$. The flow rate of new material in the hot phase is
\begin{equation}
\dot{M} = \beta(r)\dot{M}_0\bigg(\frac{r}{r_{\rm cool}}\bigg).
\end{equation}
At $r > r_{\rm d}$, $\beta(r)=1$, while for $r<r_{\rm d}$, $0 <
\beta(r) < 1$. Let $\bar{\beta}$ be the average value of $\beta$
integrated between $0 < r < r_{\rm d}$, If there is not much recycling
of material, then $\bar{\beta}$ will be close to unity. The average
mass flux of new material in the hot phase at $r < r_{\rm d}$ is then
$\bar{\beta}\dot{M}_0 r_{\rm d}/r_{\rm cool}$. Using this, and by
setting $m/\dot{m} \equiv \tau_{\rm m}$ as the average ablation
timescale of any cloud, we can write
\begin{equation}\label{eq:pimass}
\Pi_{\rm mass} = \bigg(\frac{1}{\tau_{\rm m}}\bigg)\bigg(\frac{r_{\rm
cool}}{r_{\rm d}}\bigg)\bigg(\frac{M_{\rm d}}{\bar{\beta}
\dot{M}_0}\bigg).
\end{equation}
If $\bar{\beta}$ is close to unity, then $\Pi_{\rm mass}$ is small
everywhere. Alternatively, if $\bar{\beta}$ is significantly less than
unity, $\Pi_{\rm mass}$ could be close to unity or larger. Equation
(\ref{eq:pimass}) is written in this way because we wish to express it
in terms of observable quantities, such as the total mass of cold
clouds, $M_{\rm d}$. $r_{\rm d}$ is likely to be approximately equal
to the maximum distance of clouds from the central galaxy. This
scenario is pictured schematically in figure 1.

\begin{figure}\label{fig:1}
\centering 
\includegraphics[width=8cm]{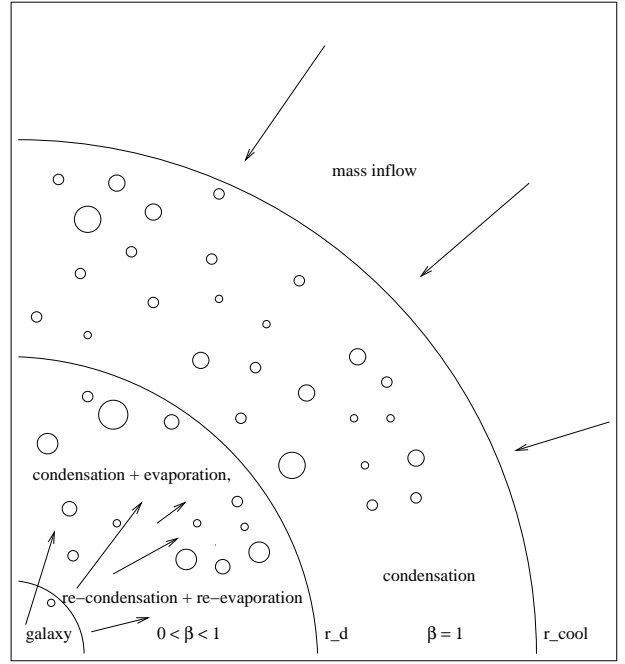}
\caption{Schematic view of the mass cycle described in the
text. Material flows inwards through $r_{\rm cool}$ and a fraction
might condense, so, for $r_{\rm d} < r < r_{\rm cool}$ we have
$\dot{M}_{\rm c}(r) < \dot{M}_{\infty}$. For $r < r_{\rm d}$
condensation and evaporation processes both occur as well as
re-condensation and re-evaporation.}
\end{figure}

\subsection{Estimated mass-coupling parameter values}

We consider the sample of seven galaxy clusters studied by
\cite{pope06}. $\dot{M}_0$, $r_{\rm cool}$ and $M_{\rm d}$ are the
mass flow rate, the cooling radius and the mass of cloud gas,
respectively. Values for them and for the mass coupling parameter
(based on the following discussion) are given in table 1.

\begin{table}
\centering
\begin{minipage}{80mm}
\caption{Parameters for the sample of galaxy clusters used in deriving
the cold gas masses and coupling parameters. Values in columns 2-3 are
taken from Pope et al. (2006) and references therein. The values in
column 4 are taken from Simionescu et al. (2007) for Virgo and from
Edge (2001) for all others. Column 5 shows the mass coupling parameter
for a region enclosing 30 kpc around the cluster centre. The values in
column 5 could be greater if $\bar{\beta} < 1$.}
\begin{tabular}{lcccc}
\hline Object & $\dot{M}_0\,(\rm M_\odot\,yr^{-1})$ & $r_{\rm cool}$ (kpc) & $M_{\rm d}\,(10^{9}\rm M_\odot)$ & $\Pi_{\rm mass}$\\ 
\hline
Virgo & 1.8 & 35 & 0.5 & 1\\ 
Perseus & 54 & 102 & 17 & 0.1\\ 
Hydra & 14 & 100 & $<$2 & $<$0.4\\ 
A2597 & 59 & 129 & 8.1 & 0.1\\
A1795 & 18 & 137 & $<$2.7 & $<$0.4\\
A2199 & 2 & 113 & $<$2.9 & $<$3\\
A478 & 150 & 150 & 4.5 & 0.06\\ 
\hline
\end{tabular}
\end{minipage}
\end{table}\label{table1}

The rate at which mass is ablated from a cloud, $\dot{m}$, might be
estimated by the rate at which momentum associated with the wind is
transferred to the cloud, divided by a speed, $c_{\rm s,a}$. This
speed would take a value anywhere between the sound speed of material
in the cloud, and the difference between the cloud speed and the
average speed of material in the filament drawn from the cloud. The
appropriate value of $c_{\rm s,a}$ depends on how effective the
viscous coupling (due to turbulence or any other mechanism) between
the cloud material and the wind material is. The rate at which mass is
lost by the cloud would then be \citep[e.g.][]{pope08}
\begin{equation}
\dot{m} = A \rho_{\rm c}\frac{(u-v)^{2}}{c_{\rm s,a}},
\end{equation}
where $A$ is the cloud cross-sectional area. The ablation timescale is
\begin{equation}\label{eq:taum}
\tau_{\rm m} \equiv \frac{m}{\dot{m}} \sim \bigg(\frac{\rho_{\rm
d}}{\rho_{\rm c}}\bigg)\bigg(\frac{c_{\rm
s,a}}{u-v}\bigg)\bigg(\frac{r_{0}}{u-v}\bigg),
\end{equation}
where $r_0$ is the initial cloud radius. We are aware of the large
uncertainties in the parameters, but consider the following estimates
suggestive: $r_0 \sim 50\,$pc, $\rho_{\rm d}/\rho_{\rm c} \sim
10^{6}$, $u-v \sim 10^{8}\,{\rm cm\,s^{-1}}$, $c_{\rm s,a} =
10^{6}\,{\rm cm\,s^{-1}}$ and $\tau_{\rm m} \sim 10^{9}$ yrs. Finally,
we take the outer radius of cold material to be $r_{\rm d}\sim
30\,{\rm kpc}$. The mass coupling parameters calculated from equation
(\ref{eq:pimass}) are shown in column 5 of table 1. These values
suggest that in the central regions at least, the mass coupling may be
a two-way process in some clusters, and mass-loss from clouds may be
significant in comparison with the mass deposited within the central
regions due to the actual flow of new material. Note that the values
in table 1 are based on the highly conservative assumption that
$\bar{\beta}=1$. If $\bar{\beta} < 1$, then $\Pi_{\rm mass}$ could be
significantly higher. A caveat is that the estimated value of
$\Pi_{\rm mass}$ also depends on the actual mass loss rates from the
clouds, which are very uncertain. Nevertheless, it seems that the mass
coupling is probably an important consideration that has been omitted
in numerical simulations of the ICM to date.

The mass coupling parameter may also be interpreted in a slightly
different way. The inflow of material, due to cooling, will deposit
material onto the central galaxy of the cluster. Current thinking
suggests that AGN activity lifts a fraction of this back into the ICM
- this allows $\Pi_{\rm mass}$ to exceed unity. We can estimate the
mass of cold gas in each cluster, by setting $\Pi_{\rm mass}=1$, and
rearranging for $M_{\rm d} = \bar{\beta}\tau_{\rm m}(30\,{\rm
kpc}/r_{\rm cool})\dot{M}_{0}$. These values are given in table 2, for
$\bar{\beta}=1$. The estimated mass of cold material will be less if
the ablation timescale of the clouds is smaller than assumed.

\begin{table}\label{table2}
\begin{minipage}{80mm}
\caption{Cold gas masses within 30 kpc calculated by setting $\Pi_{\rm
mass}=1$, $\bar{\beta}=1$ and assuming $\tau_{\rm m} = 10^{9}$yrs. The
observed mass for Virgo is taken from Simionescu et al.(2007), all
others are taken from Edge (2001).}
\begin{tabular}{lcc}
\hline Object & $M_{\rm d}/10^{9}\,{\rm M_{\odot}}$ & Observed mass/$10^{9}\,{\rm M_{\odot}}$\\
\hline 
Virgo & 1.5 & 0.5\\ 
Perseus & 15.9 & 17\\ 
Hydra & 4.2 & $<$2\\ 
A2597 & 13.7 & 8.1\\ 
A1795 & 3.9 & $<$2.7\\ 
A2199 & 0.53 & $<$2.9\\ 
A478 & 30 & 4.5\\
\hline
\end{tabular}
\end{minipage}
\end{table}

The values in column 2 of table 2 compare relatively well with the
observed quantities of cold gas in column 3. This suggests that the
ablation timescale given by equation (\ref{eq:taum}) may be a
reasonably accurate description of the process. Furthermore, since
$\tau_{\rm m} \propto (u-v)^{-2}$, it also suggests that $u-v$ cannot
be too different, on average, to $ 10^{8}\,{\rm cm\,s^{-1}}$.The small
differences between the values may be caused by slight differences
between the assumed and actual value of $\tau_{\rm m}$, by recent AGN
outbursts which may have temporarily lifted significant quantities of
cold material, or by differences in the cloud masses.

\section{Powering the optical emission in galaxy clusters}

In clusters of galaxies, the momentum transfer and dissipation of
energy by drag into the hot gas are generally insignificant compared
to gravity and radiative cooling in the hot gas. However, energy is
still dissipated by drag in significant quantities. This is
interesting because the ratios of ${\rm H}_{2}$ near and mid-infrared
to H recombination line strengths
\citep[e.g.][]{jaffe01,jaffe05,john07} are compatible with heating
being due to dissipative processes \citep[][]{ferland}. It is possible
that energy dissipation due to drag constitutes such a process.

We have already described how CO maps show clumpy structure
\citep[e.g.][]{salome08} and that the upper limit on the sizes of
these clumps, in the Perseus cluster, is $\sim 450\,{\rm
pc}$. \cite{salome08} suggest it is possible that these clumps are
complexes of giant molecular clouds. If this is true, the limitations
of spatial resolution mean it is impossible to trace the emission from
a single cloud. Consequently, theoretical estimates of the emission
due to physical processes involving the interaction between hot and
cold phases, must be applied to a volume containing multiple clouds,
rather than to an individual cloud. The total mechanical energy
injection rate required to produce filaments of average length $l$,
with a combined mass $M_{\rm d}$ and a velocity shear along their
length of $\Delta v$, is
\begin{equation}\label{eq:edotvol}
\dot{E} = \delta M_{\rm d} \frac{\Delta v^{3}}{l},
\end{equation}
where $l$ is the typical length of the filaments and $\delta$ is a
factor that takes into account the distribution of mass in terms of
its momentum. If ${\rm d M}/{\rm d}v$ and the acceleration along the
filament are both constant, then $\delta = 1/12$.

Equation (\ref{eq:edotvol}) can be written in terms of scaled
quantities as
\begin{eqnarray}\label{eq:edotvol2}
\dot{E} \sim 10^{41}\bigg(\frac{M_{\rm d}}{10^{9}\,{\rm
M_\odot}}\bigg)\bigg(\frac{\Delta v}{300\,{\rm
km\,s^{-1}}}\bigg)^{3}\bigg(\frac{l}{15\,{\rm kpc}}\bigg)^{-1}\, {\rm
erg\,s^{-1}}.
\end{eqnarray}
Interestingly, equation (\ref{eq:edotvol}) shows that the mechanical
energy injection rate is directly proportional to the total mass of
the cold material. This gives a plausible theoretical explanation for
the possible linear correlation between molecular gas mass and optical
luminosity in figure 9 of \cite{edge01}.

We can use equation (\ref{eq:edotvol2}) to estimate typical energy
injection rates assuming that each of the parameters takes its typical
value (e.g. $\Delta v = 300\,{\rm km\,s^{-1}}$, as seems to be the
case in the Perseus cluster) and taking the mass of cold gas from
column 3 of table 2. Consider the Virgo cluster; we find that the
typical values give $\dot{E} \sim 5 \times 10^{40}\,{\rm erg\,s^{-1}}$
compared to the observed $\sim 2 \times 10^{40}{\rm erg\,s^{-1}}$. In
the Perseus cluster, equation (\ref{eq:edotvol2}) gives $\dot{E} \sim
2 \times 10^{42}\,{\rm erg\,s^{-1}}$ while the optical nebula around
NGC 1275 actually emits $4.7 \times 10^{42}{\rm erg\,s^{-1}}$ in
H$\alpha$ and [NII]. Thus, equation (\ref{eq:edotvol2}) seems to
provide a reasonable explanation for the optical luminosities of both
systems.

Of course, these estimates have not accounted for different average
filament lengths, nor the variation in relative velocity between the
hot and cold phases in the different clusters. 

We can attempt to account for different average filament lengths by
comparing the sizes of the optically-emitting regions of the
clusters. \cite{heckman} give the sizes for Virgo, Perseus, Hydra-A,
A2597 and A1795 which have optically-emitting regions of linear
diameter 14, $>$53, 13, 22 and 61 kpc, respectively.  We note that the
optically-emitting region of A1795 is dominated by a single filament,
so that the remainder of the optical emission is probably contained
within $\sim 15\,$kpc, which provides a more realistic
estimate. \cite{heckman81} also gives the size of the optical region
in Perseus as $\sim 60\,$kpc. Using these values we can estimate the
average length of a filament by assuming it is a constant fraction of
the total diameter of the optically-emitting region, for example
0.25. This gives average filament lengths of $\sim 4, 15, 3, 6\,$and
$4\,$kpc for the filaments in Virgo, Perseus, Hydra-A, A2597 and
A1795, respectively. The average filament lengths of the clusters that
we do not have diameters for (A2199 and A478) will be taken to be 5
kpc, which is representative of the sample.

For the velocity shears, we have an estimate of $\Delta v \sim
300\,{\rm km\,s^{-1}}$ along the length of the filaments in Perseus
from \cite{hatch}, but we do not have similarly detailed observations
for any other clusters in the sample. In addition, linewidths at other
frequencies do not give an indication of a single well-defined value
of $\Delta v$ in each cluster. Therefore, we will assume a canonical
value for $\Delta v$ as $300\,{\rm km\,s^{-1}}$ which we know applies
adequately in the Perseus cluster. This will then be scaled to each
cluster by multiplying it by the ratio of the cluster gravitational
potential at the maximum extent of the optically-emitting region to
that of the Perseus cluster. That is, the value of $\Delta v$ that we
will use for a given cluster will be $300\,{\rm km\,s^{-1}} \times
(\phi_{\rm cluster}/\phi_{\rm perseus})^{1/2}$. \footnote{The
square-root occurs because $v \sim \phi^{1/2}$.} Such an approach is
ad hoc, but does attempt to account for the different
environments. Henceforth we will refer to this multiplicative factor
by $f \equiv (\phi_{\rm cluster}/\phi_{\rm perseus})^{1/2}$, see table
3. The gravitational potentials are estimated from the temperature and
density profiles for each cluster, with the assumption that the gas is
in hydrostatic equilibrium. Fits to the temperature and density data
are taken from \cite{pope06} and they were numerically integrated from
the origin to obtain the gravitational potential at the edge of the
optically emitting region. It was not possible to determine a
gravitational potential for A2199 due to the power-law fits to the
data. Consequently, we assigned $f = 1$ for this object.

These scalings have been applied to the sample of clusters used in
\cite{pope06} and \cite{pope07}. The optical luminosities for this
sample are taken from \cite{crawf} and \cite{heckman}. As can be seen
from table 3, the expected energy injection rates are generally
comparable with the H$\alpha$+[NII] luminosities.

\begin{table*}
\centering
\begin{minipage}{140mm}
\caption{Mechanical energy injection rates calculated with equation
  (23) using the observed cold gas masses (in table 1). In column 2 we
  list the average filament lengths derived from the size of the
  optically-emitting region. The gravitational potential $f$ factors
  are given in column 3.  The estimated optical luminosities are shown
  in column 4, the observed optical luminosities are in column 5, the
  X-ray luminosities in column 6 are taken from Pope et al. (2006) and
  references therein, and the references are shown in column
  7. Reference 3 is Heckman et al. (1989); reference 4 is Crawford et
  al. (1999). Using the [NII] fluxes given by Crawford et al. (1999)
  we have calculated $L({\rm H}\alpha+{\rm [NII]})$ rather than
  $L({\rm H}\alpha)$. The luminosities from reference 3 are the sum of
  the H$\alpha$ and [NII] luminosities. Heckman et al.(1989) did not
  give uncertainties.}
\begin{tabular}{lcccccc}
\hline Object & $l$ & $f$ & $\dot{E}/10^{40}\,{\rm erg\,s^{-1}}$ & $L({\rm H}\alpha+{\rm [NII]})/10^{40}\,{\rm erg\,s^{-1}}$& $L_{\rm X}/10^{42}\,{\rm erg\,s^{-1}}$ & Ref\\
\hline 
Virgo & 4 & 0.8 & 9 & 1.9 & 9.8 & 3\\ 
Perseus & 15 & 1 & 200 & 470 & 670 & 3\\ 
Hydra & 3 & 0.8 & $<$51 & 15 & 250 & 3\\ 
A2597 & 6 & 0.8 & 210 & 270 & 430 & 3\\ 
A1795 & 4 & 0.4 & $<$8 & 82 & 490 & 3\\ 
A2199 & 5 & 1 & $<$90 & 5.9 & 150 & 4\\ 
A478 & 5 & 0.5 & 17 & 22.7 & 1220 & 4\\
\hline
\end{tabular}
\end{minipage}
\end{table*}\label{table5}

\begin{figure}\label{fig:2}
\centering
\includegraphics[width=8cm]{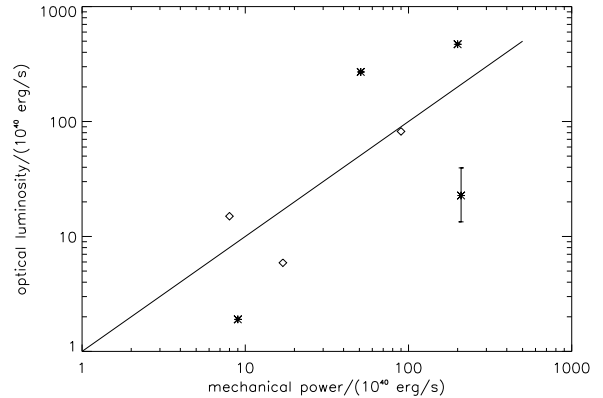}
\caption{Comparison between the mechanical energy injection rate and
H$\alpha$ + [NII] luminosity. The diamonds show the objects for which
the $\dot{E}$ is an upper limit. As a result we may expect these
points to move to the left if the mass of cold gas is less than the
upper limit. The stars denote the remainder of the sample. Ideally we
would compare the mechanical energy injection rate to $L({\rm[NII]})$,
but this is not possible since we do not have $L({\rm[NII]})$ values
for the objects in the Heckman et al. (1989) sample. The line
indicates where the optical luminosity is equal to the mechanical
energy injection rate. Note that alterations in the normalisation
could be achieved by employing a different relative velocity between
the phases. Uncertainties are shown for A478, by Crawford et
al. (1999). The other measurements are taken from Heckman et
al. (1989) who did not provide uncertainties.}
\end{figure}

Figure 2 demonstrates a relatively good match between the mechanical
energy injection rate, and the combined [NII]+ H$\alpha$ optical
luminosity. In fact, the best-fit shows that $f \times 400\,{\rm
km\,s^{-1}}$ provides a better description of the velocities. This,
particularly the scaling, agrees with the findings of
\cite{edge01}. It therefore seems possible that the dissipation of
mechanical energy may be evident in the [NII] line emission, which
traces the heating rate, while there may be an ionising source which
produces the H$\alpha$ emission. Regardless of this, the general
conclusion that optical emission increases with mechanical power is
consistent with \cite{ferland} who found that the ${\rm H}_{2}$ to H
recombination line ratios are compatible with dissipative heating.

It is worthwhile pointing out that our estimate for A1795 falls below
the observed optical luminosity. This is probably because the assumed
velocity is a conservative estimate given the current likely state of
the system. The AGN has been recently active and indeed has injected
more energy than any of the others in the sample, save for the one in
Hydra-A, which is a well-known powerful AGN, so a fast outflow would
be expected in the central regions. In addition, this system would
only requires $\Delta v \sim 300\,{\rm km\,s^{-1}}$ to account for the
optical emission, so it seems more than likely that mechanical energy
can account for the optical emission. A similar explanation also
accounts for the apparent deficit in the Perseus cluster.

The question of why some clusters are brighter than others in their
optical emission remains. The mechanical energy injection rate seems
to be proportional to the total mass of cold gas so that clusters with
more cold gas are likely to exhibit more optical emission. This leads
to the question: why is there more cold gas in some clusters than
others?  The mass of cold gas is related to the mass flow rate, which
is, in turn, related to the balance between heating and cooling in the
cluster. There may also be an additional component of material thrown
up from the central galaxy. It is important to note the strong effect
of the velocity difference between the phases. This is again likely to
be related to AGN activity, so that the optical luminosity in clusters
with AGN is likely to be larger than expected.

\subsection{Infra-red emission}

There is also significant infra-red emission from the molecular gas
near cluster centres. \cite{jaffe05} quote luminosities of $\sim
10^{42}\,{\rm erg\,s^{-1}}$ for the ${\rm H}_{2}(1-0){\rm S}(1)$ line
in A2597 and suggest that this comprises roughly 1-2\% of the total
molecular NIR emission due to ${\rm H}_2$. This would imply a total
NIR luminosity of $10^{44}\,{\rm erg\,s^{-1}}$ which greatly exceeds
the optical luminosity for A2597. \cite{jaffe97} also give
luminosities of $\sim 10^{40}\,{\rm erg\,s^{-1}}$ and $\sim
10^{41}\,{\rm erg\,s^{-1}}$ for the ${\rm H}_{2}(1-0){\rm S}(1)$ line
in A478 and Hydra-A, respectively. This implies total NIR luminosities
of $10^{42}-10^{43}\,{\rm erg\,s^{-1}}$ respectively. Such values are
comparable with the optical luminosities of the systems, although not
in the case of A2597. However, it is still feasible that the injection
of mechanical energy can also account for the NIR emission. To obtain
a mechanical energy injection rate of $\sim 10^{44}\,{\rm
erg\,s^{-1}}$ in A2597 requires $\Delta v \sim 700-800\,{\rm
km\,s^{-1}}$. Such values are comparatively large, but probably close
to a Mach number of unity for the hot gas in the cluster centre and is
therefore perfectly permissible. Furthermore, A2597 shows evidence for
a recent, moderately powerful AGN outburst \citep{birzan}; therefore,
transonic velocities close to the cluster centre are to be
expected. The same is true for Hydra-A, which also shows strong NIR
emission and currently hosts a strong AGN outburst.

\section{Discussion and summary}

The effect of mass injection into a diffuse flow in a gravitational
potential was investigated by \cite{pittard04} using 1D hydrodynamical
simulations. Regularly spaced shock-like structures were obtained
under some conditions, which could provide an alternative explanation
for the structures generally believed to be AGN-driven shocks. Other
simulations revealed a differential luminosity function which
resembled those deduced from observations
\citep[e.g.][]{peterson03}. However, along with these successes are a
number of areas for concern. Firstly, the amount of material that was
added was not related to the mass deposition rate (expected from
cooling), and eventually dominated the total gas mass in the
system. However, this is less of a problem if the flows are
fountain-like. Secondly, the resulting X-ray luminosity was far too
centrally peaked. This may be related to the form of the gravitational
potential used, and it would be worth revisting the model, to see what
the effect of a realistic cluster potential, and realistic mass
transfer rates, would have. It may also be necessary to include
thermal conduction as well, to see whether these structures would be
smoothed out by the same process that evaporates the clouds.

Our analysis in the current work indicates that the rate of mechanical
energy injection is comparable with the optical luminosity in the
sample. Since the [NII] emission is a tracer of the heating rate of
the nebular gas, it seems possible that the observed emission is
indeed powered by mechanical energy injection. Furthermore, given that
the energy dissipation rate is proportional to $(u-v)^{3}$ it is not
surprising that optical emission occurs frequently in cool-core
clusters, since AGN are frequently found in such environments. That is
not to say that radiation from the AGN is responsible for the
emission, but that AGN-induced bulk motions in the ICM are likely to
cause significant interactions with the cold material in the central
regions of clusters which may be evident through the optical
emission. It may also be revealing that optical emission in clusters
with AGN is generally more luminous than would be predicted. It is
also possible that a component of the ${\rm H}_\alpha$ is due to the
same process, though photoionisation also plays a role.  As we have
already stated, it is also possible that mechanical energy also powers
the NIR emission. Again, prominent NIR emission seems to occur in
systems with powerful AGN. As a result, it is possible that the
presence of cold material may automatically lead to physical processes
which power the observed optical and infra-red emission. In this
picture the injection of energy will necessarily be relatively
constant and distributed, in accordance with observational
constraints.

This work suggests that there are plausible theoretical reasons which
could explain the rough proportionality between the total optical
luminosity and the total mass of cold material observed by
\cite{edge01}. The total cold mass is likely to depend on the
heating/cooling balance in the cluster centres, thus providing us
important information on the deviation from equilibrium between these
processes.

Our numerical simulations \citep[][]{pope08} indicate that, if formed
by outflows, filaments can only be formed if there is a sufficiently
fast wind ($>{\rm few} \times 10^{7}\,{\rm cm\,s^{-1}}$). Without this
wind, any cold, dense material stripped from a cloud would sink
towards the cluster centre with the cloud, rather than forming a long
tail behind the cloud. Thus, a slower wind may lead to amorphous
optical emission, without filaments. Consequently, the morphology of
the optically emitting region is likely to be strongly dependent on
the magnitude of the relative motion between the ICM and the cold
phase.

To summarise, we have two main points:

1) There is enough cold gas within the central 30 kpc, or so, of most
   galaxy clusters for its evaporation to contribute significantly to
   the density of the hot gas in this region. Therefore, to obtain a
   more complete understanding of the behaviour of the ICM this effect
   should be included in future numerical simulations.

2) In general, it seems that the mechanical energy or momentum
   injection will not affect the thermal structure of the hot phase,
   but it may account for, at least, the [NII] component of the
   optical emission of the filaments and the NIR emission. Under
   exceptional circumstances this process could also alter the flow
   dynamics of the ICM and account for some heating in the central
   regions of clusters.

\section{Acknowledgements}

We wish to thank Nina Hatch and Paul Nulsen for informative
discussions and the referee, Walter Jaffe, for very helpful comments
that improved this work. JMP thanks the Royal Society for
funding. This work was funded in part by STFC.

\bibliography{database} \bibliographystyle{mn2e}

\label{lastpage} 

\end{document}